# A Multi-Case Study of Agile Requirements Engineering and the Use of Test Cases as Requirements

Elizabeth Bjarnason[1], Michael Unterkalmsteiner[2], Markus Borg[1], Emelie Engström[1]

[1]Lund University
SE-221 00 Lund, Sweden
FirstName.LastName@cs.lth.se
[2]Blekinge Institute of Technology
SE-371 79 Karlskrona, Sweden
mun@bth.se

**Abstract.**
   [*Context*] It is an enigma that agile projects can succeed 'without requirements' when weak requirements engineering is a known cause for project failures. While agile development projects often manage well without extensive requirements test cases are commonly viewed as requirements and detailed requirements are documented as test cases.
   [*Objective*] We have investigated this agile practice of using test cases as requirements to understand how test cases can support the main requirements activities, and how this practice varies.
   [*Method*] We performed an iterative case study at three companies and collected data through 14 interviews and 2 focus groups.
   [*Results*] The use of test cases as requirements poses both benefits and challenges when eliciting, validating, verifying, and managing requirements, and when used as a documented agreement. We have identified five variants of the test-cases-as-requirements practice, namely de facto, behaviour-driven, story-test driven, stand-alone strict and stand-alone manual for which the application of the practice varies concerning the time frame of requirements documentation, the requirements format, the extent to which the test cases are a machine executable specification and the use of tools which provide specific support for the practice of using test cases as requirements.
   [*Conclusions*] The findings provide empirical insight into how agile development projects manage and communicate requirements. The identified variants of the practice of using test cases as requirements can be used to perform in-depth investigations into agile requirements engineering. Practitioners can use the provided recommendations as a guide in designing and improving their agile requirements practices based on project characteristics such as number of stakeholders and rate of change.

## 1   Introduction

Agile development methods strive to be responsive to changing requirements by integrating requirements, design, implementation and testing processes (Sommerville 2005, Layman 2006). Face-to-face communication is prioritised over written requirements documentation and customers are expected to convey their needs directly to the developers (Beck 2001, Ramesh 2010). However, weak customer communication in combination with minimal documentation is reported to cause problems in customer participation and with scaling and evolving software for agile projects (Ramesh 2010).

   Requirements specifications are used for different purposes and support the main requirements activities of eliciting and validating stakeholders' requirements, software verification, tracing and managing requirements, and for contractual purposes by documenting customer agreements. We use the set of main requirements activities for which the requirements specification plays a role as identified by Lauesen (2002). Requirements are used to communicate with stakeholders, to drive design and testing, and to serve as a reference for project managers and in the evolution of the system (Davis 2005). Due to the central role of requirements in coordinating software development, there exists a plethora of research on requirements documentation with varying degrees of formality depending on its intended use. This spans from formal requirements specifications (Lamsveerde 2000) and requirements models (Pohl 2010), over templates (Mavin 2010) to user stories (Cohn 2004), and natural language specifications. Formal requirements specifications can be automatically checked for consistency (Heitmeyer 1996) and used to derive other artefacts, e.g. software designs (Dromey 2003) or test cases (Miller 2012). Less formal requirements documentation is driven by heuristics and best practices for achieving high quality requirements (Davis 1993).

The coordination of evolving requirements poses a challenge in aligning these with later development activities including testing (Uusitalo 2008, Bjarnason 2014). In a previous study we identified the practice of documenting detailed requirements as test cases, i.e. using test cases as requirements (TCR) as one of several industrial practices that addresses this challenge (Bjarnason 2014). In this paper, we investigate TCR further for the three case companies (of six) from our previous study that explicitly mentioned this practice. In this paper we discuss *how TCR may support requirements engineering* (RE) by investigating the following:

**RQ1** How can test cases of the TCR practice support the main requirements activities, i.e. elicitation and validation, verification, managing changes, and for contractual purposes (Lauesen 2002) and thus fill the role of a requirements specification? In particular, what are the benefits and the challenges of this approach?

**RQ2** What variations are there in applying the TCR practice?

Intermediate results for these research questions can be found in (Bjarnason 2015a) for the same three case companies based on 12 interviews from a previous study. For this paper, additional data was collected through two focus group sessions and two (new) interviews. In particular, we followed up on two of the three previously reported cases, one of which was a (previously) planned implementation of TCR. For this paper, further cross-case analysis and synthesis of the full set of data has been performed. We report two additional variants of the practice (*stand-alone strict* and *stand-alone manual*), and also depict the requirements flows for each case company thereby visualizing the application of TCR.

The rest of this paper is organized as follows. Section 2 describes background and related work. Section 3 presents the case companies and Section 4 the applied research method. The results are reported in Section 5 and discussed in Section 6 where the research questions are answered. Finally, the paper is concluded in Section 7.

## 2  Background and Related Work

While requirements engineering (RE) and testing are traditionally viewed as two separate processes, the TCR practice studied in this paper is an example of a practice where the activities of these two processes are performed concurrently (Lawson 1994). The work presented in this paper stems from research into the coordination and alignment of RE and testing (RET) for software development in general including traditional development. In addition, our work relates to the agile approach of integrating the RE activities with those of testing. We will now describe these two related areas of research.

### 2.1  Requirements Engineering and Test (RET) Alignment

Coordinating and aligning RE and testing is a challenge within software development projects. This challenge relates to a range of issues including organization, process, people, tools, requirements changes, traceability and measurements (Bjarnason 2014, Sabaliauskaite 2010). Alignment methods have been studied from the perspective of linking information between people and/or documentation using mechanisms of varying formalism and complexity (Unterkalmsteiner 2014). Industrial practices in this area include traceability, model-based approaches and increased communication, e.g. by involving testers in requirement reviews (Bjarnason 2014, Uusitalo 2008). Similarly, in requirements-driven collaboration there is close communication between requirements and testing roles; key roles which when absent cause disruptions within the development team (Marczak 2011).

We previously investigated RET alignment through a large interview study at six development companies (Bjarnason 2014). The results include 10 main challenges and 10 categories of practices. Examples of RET challenges include aligning goals, requirements specification quality, maintaining alignment during changes, and outsourcing testing. The main categories of RET practices include change management, tracing, tools, and metrics. The RET study identified four high-level factors that affect RET alignment. These factors are the human aspects of development, the quality of requirements, the size of the development, and the incentives for implementing alignment practices. The human side of development including communication and coordination is vital for alignment in general, so also between requirements engineers and testers. Further, the quality and accuracy of requirements is a crucial starting point for testing the produced software in-line with the agreed requirements. In addition, the size of the development organisation and its projects is a key variation factor that affects which challenges that are faced and which tools and practices that are suitable. Finally, the incentive for applying practices such as good requirements documentation and tracing vary. For companies with safety-critical development this incentive is externally motivated, while the motivation is purely internal for non-safety critical cases. This internal motivation for RET practices is often weak due to low awareness of the cost vs. benefit of RET alignment.

## 2.2 The Agile Approach of Integrating Requirements Engineering with Testing

In agile development requirements and tests can be seen as two sides of the same coin. Martin and Melnik (2008) hypothesize that as the formality of specifications increases, requirements and tests become indistinguishable. This principle is taken to the extreme by unit tests (Whittaker 2000) where requirements are formalized in executable code. Practitioners report using unit tests as a technical specification that evolves with the implementation (Runeson 2006). However, unit tests may be too technical for customers and thereby lack the important attribute of being understandable to all relevant stakeholders.

Acceptance tests are used to show customers that the system fulfils the requirements (Hsia 1997). However, developing acceptance tests from requirements specifications is a subjective process that does not guarantee that all requirements are covered (Hsia 1997). This is further complicated by requirements documentation rarely being updated (Lethbridge 2003), leading to potentially outdated acceptance tests. In agile development, automated acceptance tests drive the implementation and address these issues by documenting requirements and expected outcomes in an executable format (Ramesh 2010, Haugset 2008). This agile practice is known, among others, as customer tests, scenario tests, executable/automated acceptance tests, behaviour-driven and story-test-driven development (Park 2010).

Some organisations view and use the automated acceptance tests as requirements thereby fully integrating these two artefacts (Martin 2008). Automated acceptance tests are used to determine if the system is acceptable from a customer perspective and provide the basis for customer discussions, thus reducing the risk of building the wrong system. However, more technical communication might be needed which requires technical insight of the customer. Melnik et al. (2006) found that customers in partnership with software engineers could communicate and validate business requirements through automated acceptance tests, although there is an initial learning curve.

The conceptual difficulty of specifying tests before implementation (Causevic 2011, George 2004, Janzen 2007) led to the conception of behaviour-driven development (North 2006). This approach incorporates aspects of requirements analysis, requirements documentation and communication, and automated acceptance testing. The behaviour of a system is defined in a domain-specific language; a common language that reduces ambiguities and misunderstandings. This is further enhanced by including terms from the business domain in the domain-specific language. Solis and Wang (2011) reviewed the available literature and a number of tools for behaviour-driven development in 2011. They found that the area was still under development and that the domain-specific languages supported by the tools limited the requirements expressiveness.

Haugset and Hanssen studied acceptance test driven development (ATDD) as an RE practice and report on its benefits and risks (Haugset 2008). Our work extends on this by also investigating companies that use the TCR practice without applying ATDD principles.

## 3 Case Companies

The three case companies all develop software using an agile development model. However, a number of other factors vary between the companies. These factors are summarised in Table 1.

Table 1. Overview of the case companies.

| Company | A | B | C |
|---|---|---|---|
| Type of company | Software development, embedded products | Consulting | Software development, embedded products |
| #employees in software development | 250 (125-150 in 2009) | 135 | 1,000 |
| #employees in typical project | 10 | Mostly 4-10, but varies greatly | Product projects: 400-500 Feature projects: 3-15 |
| Distributed | No | No | Yes |
| Domain / System type | Computer networking equipment | Advisory/technical services, application management | Telecom |
| Source of requirement | Market driven | Bespoke | Bespoke, market driven |
| Main quality focus | Availability, performance, security | Depends on customer focus | Performance, stability |
| Certification | Not software related | No | ISO9001 |
| Process Model | Agile | Agile in variants | Agile with gate decisions |
| Project duration | 6-18 months | No typical project | Previously: 2 years |
| #requirements in typical project | 100 (20-30 pages HTML) | No typical project | 2,000-3,000 |
| #test cases in typical | ~1,000 test cases | No typical project | 15,000 |
| project | | | |
| **Product Lines** | Yes | No | Yes |
| **Open Source SW Used** | Yes | Yes | Yes |

### 3.1 Company A

Company A develops network equipment consisting of hardware and software. The software development unit covered by the interview study had around 150 employees in 2009 and grew to 250 employees in 2015. The company is relatively young but has been growing fast during the past few years. A typical software project has a lead time of 6-18 months, around 10 co-located software developers and approximately 100 requirements and 1,000 system test cases. The company has a mature agile development process with a strong focus on testing. A market-driven requirements engineering process is applied. The quality focus for the software is on availability, performance and security. Furthermore, the company applies a product-line approach and uses open-source software in their development of closed-source products.

The participants from Company A consist of three interviewees from a previous study and five focus group participants. The interviewees were a product manager, a project manager, and a tester, see Table 2. The focus group participants consisted of two product managers, a systems architect, a technical project manager and a tester, see Table 3.

**Table 2.** Interviewees per company selected from a previous study: roles and experience. For Company B, software developers also perform RE and testing tasks.

*Legend: Experience in role noted as S(enior) = more than 3 years, or J(unior) = up to 3 years. Only interviewees mentioning the TCR practice were included in this study, these are marked with **bold**.*

| Role | A | B | C |
|---|---|---|---|
| Requirements engineer | | | **C1:S**, C6:S, C7:S |
| Systems architect | | | C4:S |
| Software developer | | B1:J, B2:S, B3:S | C13:S |
| Test engineer | **A2:S** | | C9:S, **C10:S**, C11:J, C12:S, C14:S |
| Project manager | **A1:J** | | **C3:J, C8:S** |
| Product manager | **A3:S** | | |
| Process manager | | | **C2:J, C5:S**, C15:J |

### 3.2 Company B

Company B is a consultancy firm that provides technical services to projects that vary in size and duration. Most projects consist of one development team of 4-10 people located at the customer site. The company has applied agile development practices for more than a decade and are active in the agile community. The requirements are defined by a customer (bespoke).

The three consultants that were interviewed at Company B can mainly be characterised as software developers, see Table 2. However, they all typically take on a multitude of roles within a project and are involved throughout the entire lifecycle.

**Table 3.** Participants in focus group for Company A. Participant ids A21.n where A21 denotes the focus group session and n the participant in that session.

| | Roles | | Experience (years) | | | |
|---|---|---|---|---|---|---|
| | Current | Previous | Current role | Previous role | At company | Total |
| A21.1 | Product manager | Systems architect | 1 | 7 | 18 | 20 |
| A21.2 | Product manager | Business manager | 10 | 2 | 13 | 30 |
| A21.3 | Systems architect | Software developer | 5 | 10 | 19 | 25 |
| A21.4 | Technical project manager | Software developer | 2 | 10 | 5 | 12 |
| A21.5 | Tester | Tester | 5 | 3 | 8 | 10 |

### 3.3 Company C

Company C develops software for embedded products in the telecommunications domain. The software development unit investigated in this study, consists of 1,000 people. At the time of the initial interviews, the company was transitioning from a waterfall process to an agile process. Product projects typically run over 12-18 months and include around 400-500 people, while software features are developed in smaller sub-projects consisting of around 3-15 people. The projects handle a combination of bespoke and market-driven requirements. Including the product-line requirements, they handle a very complex and large set of requirements.

Six (of fifteen) interviews from a previous study were relevant to include in this study. These interviews were with one requirements engineer, two project managers, two process managers and one tester, see Table 2. Additional interviews were held with one systems architect and one account manager at Company C, see Table 4. Furthermore, six software architects participated in a focus group session. These software architects represent six software development teams, or feature projects, that develop software ranging from high-level user applications to software utility functions. The user applications aim to release new software updates every 6 weeks while the utility functions are more tightly coupled to the hardware and product releases, which have a 6-monthly release cycle. These feature development projects range in size from 3 to 15 project members.

Table 4. Characterisation of participants in focus group (C21) and interviews (C22, C23) for Company C in iteration 2.

| Data collection entity | | Roles | Experience (years) | |
|---|---|---|---|---|
| | | | Current role | Total |
| C21 | Focus group | 6 software architects | 2-10 | 5-20 |
| C22 | Interview | Systems architect | 9 | 20 |
| C23 | Interview | Account manager | 12 | 30 |

# 4 Method

The motivation for this research hails from our previous study on practices used to align RE and Testing (RET) (Bjarnason 2014), one of which is the use of test cases as detailed requirements documentation. In order to gain an in-depth understanding of this practice, we performed a case study in two iterations using a flexible exploratory case study design and process (Runeson 2012). This approach allowed us to explore and compare variations of the practice found in the three case companies. In the first iteration we analysed interview data collected during 2009/2010 as part of our previous study where a wider set of RET practices were identified. The purpose of the first iteration was to gain an understanding of how the TCR practice is applied in industry; its benefits and challenges. The second iteration was performed based on the outcome of the first iteration, in particular to follow up and further investigate the situation for the two case companies for which the practice was less mature (Companies A and C) and to complement the (more extensive) data available for Company B. We collected additional data through focus groups and interviews for case companies A and C during Spring 2015 to see if the practice had changed or matured, and to investigate the implementation of TCR in more depth. For each of the two iterations, a case study process (Runeson 2012) was applied consisting of four stages: 1) *Definition & Planning*, 2) *Data selection/collection*, 3) *Data analysis* and 4) *Reporting*. An overview of the method for each iteration is shown in Table 5.

Table 5. Overview of the applied case study process for each of the iterations.

| *First iteration (data collected 2009/10)* | *Second iteration (data collected 2015)* |
|---|---|
| **Definition and Planning** | |
| RQ1. How can test cases of the TCR practice support the main requirements activities, i.e. elicitation and validation, verification, managing changes, and for contractual purposes (Lauesen 2002) and thus fulfil the role of a requirements specification? In particular, what are the benefits and challenges of this approach? | |
| RQ2. (1st version) Why and how is the TCR practice applied? | RQ2. What variations are there in applying TCR in practice? |
| **Data Selection / Collection** | |
| Selected relevant interview data from previous study (Bjarnason 2013) for Companies A, B and C. In total 12 semi-structured interviews. | |
| | Collected additional data for Companies A and C through 2 focus group sessions and 2 semi-structured interviews. |
| **Data Analysis** | |
| Word-by-word transcripts | Transcriptions |
| Descriptive coding and clustering into benefits and challenges | Semi-exploratory coding based on Lauesen's set of main requirements activities (RQ1), and roles and artefacts in the |

|  | requirements flow. |
| --- | --- |
| Triangulation applied by (at least) one other researcher reviewing each transcription and coding. | |
| Transcripts reviewed by interviewee. | Summary of transcript reviewed by participants/interviewee. |
|  | Cross-cases analysis |
| **Reporting** | |
| In Bjarnason 2015a | In this paper |

## 4.1 First Iteration: Initial Exploration

In the first iteration we analysed existing interview data from our previous (wider) study of industrial practices in aligning RE and testing (Bjarnason 2014). For the current paper, we analysed the interview data relevant to the TCR practice in more detail. This iteration consisted of defining research questions, selecting data, analysing this data and reporting of the results in (Bjarnason 2015a). A description of the applied method for the initial study can be found in Runeson 2012, chapter 14 and appendix C. In this paper, we describe the steps taken for this follow-on study.

### 4.1.1 Definition of Research Questions and Planning

Since we are interested in how agile development can be successful 'without requirements' we selected to investigate the practice of using test cases as requirements. We formulated the research questions, (RQ1) How can test cases of the TCR practice support the main requirements activities? and (RQ2, 1st version) Why and how is the TCR practice applied?

### 4.1.2 Data Selection

We selected to use word-by-word transcriptions from our previous study of RE-Testing coordination (RET), which also included agile processes. The research questions of this iteration are within the broader scope of the previous study on RET (Bjarnason 2014), i.e. the previously collected data contains information relevant to these questions. Furthermore, the semi-structured interviews provided rich material since the interviewees could freely describe how practices (of which TCR was one) supporting RET were applied including benefits and challenges. Data selection was facilitated by the rigorous coding performed in the previous study. We selected the interview parts coded for the TCR practice. In addition, the transcripts were searched for key terms such as 'acceptance test' and 'specification' to further ensure that all relevant interview parts were selected to be included in this study.

### 4.1.3 Data Analysis

The analysis of the selected interview data was performed in two steps. First the transcripts were descriptively coded. These codes were then categorised into benefits and challenges and reported per case company in (Bjarnason 2015a). The analysis was performed by the first author. The results were validated independently by the third author. The third author analysed and interpreted a fine-grained grouping of the interview data produced in the previous RET study (Bjarnason 2014), and compared this against the results obtained by the first researcher. No conflicting differences were found.

## 4.2 Second Iteration: Complementary and Focused Investigation

The aim of the second iteration was to complement the insights of the first iteration and to gain a deeper understanding of the variations of the TCR practice. For this we collected new empirical data focused specifically on the TCR practice and also performed further analysis of the full set of empirical data.

### 4.2.1 Case Study Design: Definition of Research Questions and Planning

The authors defined and planned the second iteration of this case study over a period of one month. The two research questions of the first iteration were selected to be included also for this iteration, however RQ2 was rephrased to more specifically focus on variations in applying TCR in practice (RQ2), i.e. how the practice is applied in different contexts. Thus, motivation (how) for applying the practice was removed for the final version of RQ2.

We decided to gather new empirical data to complement our set of data for the two companies where the practice was previously found to be weaker and for which there was less data in the first iteration, namely Company A and C. For Company B there was already a substantial amount of relevant data from the first iteration. Thus, we decided to investigate the practice further at Company A and C. This also allowed us to follow-up on the implementation of TCR that was planned as part of the agile transition at Company C during the first iteration.

**A data collection protocol** was defined based on the research questions. This protocol contains questions on how TCR was applied or could be applied to the company's current requirements information flow and how each main requirements activity was fulfilled by TCR. The data collection protocol was designed and agreed jointly by the researchers, see Appendix.

**The data collection method** used was mainly focus group sessions due to the interactive nature of agile RE, which involves collaboration between many different roles. A focus group allows for eliciting a holistic view of the situation from multiple perspectives, as well as, providing value in itself to the participants through group reflection and learning (Robson 2002). Semi-structured interviews using the same data collection protocol were used as a compliment to the focus group sessions. The data collection is further described below.

**Sampling of participants** was done with the aim of covering all roles throughout the life cycle from customer to testing. However, due to practical reasons (limited time frame and participant availability) this was combined with convenience sampling and adaptions to the data collection method. A characterization of the participants is provided in Table 3 and Table 4.

### 4.2.2 Data Collection

In the second iteration data was collected from Company A through one focus group session and from Company C through one focus group session and two semi-structured interviews. The interviews were used to investigate two additional applications of TCR within Company C, mentioned by company representatives during the planning phase, namely API implementation and self-certification of customer requirements.

Each focus group session was attended by two researchers; one moderator who lead the discussion and one note-taker who asked follow-up questions when needed. The interviews were performed by one researcher. The focus group sessions and the interviews were audio recorded after permission was granted by the participants.

The data collection protocol (see Appendix), was used as a guide during the focus group sessions and the interviews. In the introduction part the participants were informed of the aim of the study, how it would be performed, and confidentiality. The participants were then asked to introduce themselves and their role. The overall information flow between requirements and testing including the used artefacts and involved roles was then explored. This was followed by semi-structured discussions on how test cases can fulfil the various main requirements activities. The meeting was then closed with a summary discussion around strengths and weaknesses with the practice and possible improvements.

The duration of the data collection sessions varied between 30 minutes for the interviews through 45 minutes for the focus group at Company C and 2 hours for the focus group at Company A. The reason for the short focus group session at Company C was limited availability of the participants, which also meant that only project architects were present at this session. The longer meeting duration for the focus group at Company A allowed for more individual reflection time and jointly constructing an overview of the requirements flow using post-it notes produced individually by the participants and then presented and jointly discussed.

### 4.2.3 Data Analysis

The audio recordings of the focus group sessions and the interviews were transcribed. The focus group sessions were transcribed by the note-taking researcher and checked by the moderating researcher. The interviews were transcribed by the interviewing researcher.

The transcripts were coded using a semi-exploratory coding approach using initial codes based on the research questions, e.g. main requirements activities, involved artefacts and roles. For each transcript the coding was performed by one researcher and reviewed by another. These codes for artefacts and roles were used to construct the diagrams depicting the flow of requirements information for each case, see Figure 1, 2 and 3.

Finally, a cross-case analysis was performed per research question through comparative analysis of the full set of empirical data using the existing coding to locate parts of the transcripts concerning similar topics. This analysis was performed by comparing the use of the TCR practice between the companies, thereby identifying common themes and thus synthesizing the findings of our multi-case study. For example, for RQ1 while analysing how the TCR was applied, similar benefits and challenges were identified for different companies. This enabled us to compare and contrast the pros and cons of the practice between cases. Similarly for RQ2, a set of varying facets of the practice emerged while analysing the variations between the case companies concerning how they apply the TCR practice.

Full traceability was maintained from the transcript through the coding and into the reporting, i.e. this article. This is denoted by references from the results to the empirical data; the individual transcripts and chunk id. These traces were also used for reviewing and validating the reported results against the data.

#### 4.2.4 Reporting

As a first step of reporting the results of the newly collected data, a summary was produced per transcript structured per main requirements activity. This summary was sent to the participants to review and contribute with possible additional information or corrections. The results from both from both iterations were merged to providing a uniform story of TCR for the three case companies.

In order to convey the rich data concerning the requirements information flows derived from the focus group sessions we decided to visualise these flows using the FLOW notation by Stapel and Schneider (2012). The reason for choosing this notation is that it is designed to convey communication situations and patterns, rather than a formal process. The notation conveys the use of both documented (solid) information and un-documented (fluid) information, of which the latter in particular plays an important role within agile development. Information storage is represented by a document symbol for solid information, and a smiley for fluid, or un-documented information, see Table 6. Fluid information storage is typically a role or an individual, we also use it to denote a function, e.g. development team to indicate all the roles within a development team. The flow of fluid information is denoted by a dashed line, while a solid line is used to represent communication of solid information.

**Table 6.** The notation used to illustrate the requirements flows for the case companies (see Section 5), based on the FLOW notation (Stapel and Schneider 2012).

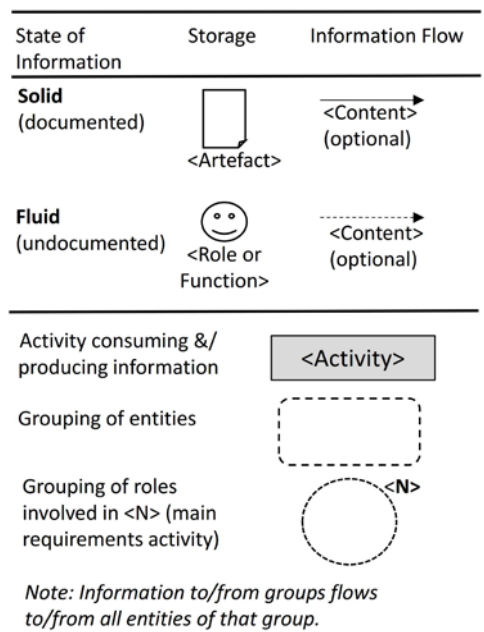

In this paper we use flow diagrams to present information flow for each of the main requirements activities (e.g. elicitation, validation) and high-light the relevant entities by encircling the people (fluid information storages) involved in each requirements activity with a dashed lines (an addition to standard FLOW notation). The information that flows to and from a circle is shared by all the fluid information storages within the circle. If a specific storage (role or function) is involved in a flow, e.g. produces an artefact, this is denoted by a line to/from the icon representing that specific storage rather than to the circle.

Most fluid information storages and some artefacts are duplicated and thus appear several times in the same figure. This is to provide an overview of the requirements flow for each main requirements activities. Flows between requirements activities is primarily denoted by connections to common artefacts (solid information storages).

Furthermore, timing aspects are not captured by the notation, e.g. different versions of an artefact or the order of activities.

## 5    Results: Applications of TCR in Practice

All of the investigated companies apply the TCR practice, however the context, extent and maturity of the practice varies. Based on the data collected for each company, we will now describe how TCR is involved in eliciting, validating, verifying and managing requirements, and in documenting the agreement with the customer. For each case company the application of TCR for software development is described. In addition, for Company C the TCR practice was also found to be used for API development. The requirements flow for each case context is depicted in Figure 1 – Figure 4 based on the FLOW notation (Stapel and Schneider 2012). The notation shows the flow of non-documented (fluid) and artefact-based (solid) requirements communication, see Section 4.2.4. The description of the results contain two additional types of references. One denoted [Nn[1]] that provides traces from the results to the empirical data, and one denoted, e.g. (EB3[2]), (MC1) that refers to benefits and challenges found to be common for the case companies (these are discussed in Section 6.1).

### 5.1    Company A: TCR as a De Facto Practice

The test cases have become the de facto requirements at Company A since the most extensive set of requirements are documented as test cases as part of the testing process [A21], i.e. once the requirements details have been implemented [A21].The test cases consist of a combination of unit, function and system level test cases produced by the system testers, the developers and the system architects [A21]. These test cases are kept updated [A21] and are the most accurate requirements documentation [A21]. Other requirements documentation, e.g. the product requirements specification, is not maintained past the initial implementation stage [A21]. The same basic process was identified in both iterations of our study. An overview of the requirements flow is shown in Figure 1.

The test cases provide valuable requirements information [A21] and are used to understand product details [A21] by product managers with technical competence [A21] and roles within development, such as system architects and developers [A21]. The product manager said: 'I read test cases … because that is what we can say actually works' [A21]. The test cases are also used as a source of requirements information in selecting the test scope for integration of new features [A21] and for analysing the impact of changes [A21].

At times the test results are used in the requirements communication, in particular concerning unspecified quality requirements. Senior engineers and developers then view the test results, and if they agree to the test outcome, it becomes a requirement that is included in the test cases [A21].

This co-located organisation finds TCR to be an efficient way of managing requirements [A1] that decreases the number of documents to maintain and thus the overhead of coordinating RE and testing [A21]. The lack of formal requirements is only experienced as a problem when test case validity is questioned, e.g. when a test fails. The requirement's motivation and priority, which is not documented, then have to be re-investigated (MC2). The navigation of the test case 'requirements' is simplified by similar structuring in test cases and source code [A1]. A product manager suggested that the descriptions noted in the test cases of the verified use case and (sometimes) its motivation could be extracted and thereby be made more readily available to the development projects (CC1) [A21]. This is useful requirements information and could possibly be extended and replace other artefacts [A21].

Implementation is managed in a Scrum-like way with a backlog prioritized by the product manager from which the development project select work items based on priority, cost and feasibility [A21]. The backlog contains new requirements, issues and change requests, which are all handled in a similar way.

---

[1] Nn refers to interview transcript Nn, where N is company A, B or C, and n is the transcript number, see Section 3.

[2] EBn, ECn, VBn, VCn, TBn, TCn, MBn, MCn, CBn, and CCn denote RE area, Benefit or Challenge, and sequence number, see Table 7.

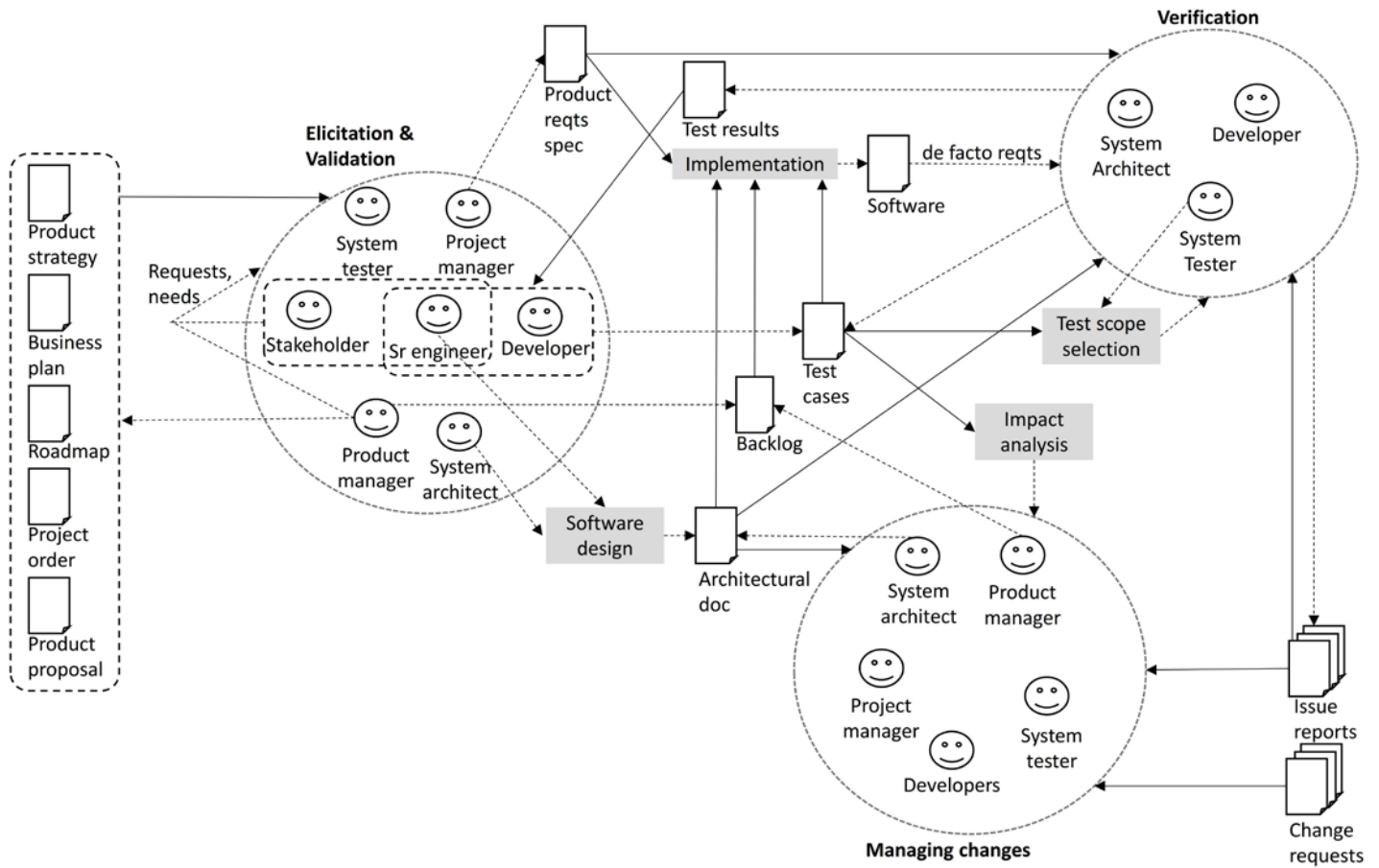

**Figure. 1.** Overview of requirements flow for Company A's product development. Notation described in Section 4.2.4.

### 5.1.1 Elicitation and Validation: Company A

Early and very high-level requirements are documented by the product manager in business plan, product strategy and roadmap documents [A21], who also creates a project order and a product proposal [A21]. The requirements are then detailed in close collaboration between project roles (EB1), i.e. integrated with software design [A21], implementation, and testing [A21]. This is believed to be due to having a strong testing competence within the company, but no RE-specific role or competence (EC3) [A2]. Instead, the product manager acts as the proxy for market and customers with whom the development-near roles have no direct contact (EC2). During development, direct communication between the product manager and the development-near roles is the main requirements communication channel [A21] complimented by a high-level product requirements specification produced by the project manager [A21] and architectural artefacts created by the system architect during software design [A21]. These architectural documents include an overview of the software architecture and functional descriptions of the product, and represent a contract between the product manager (who orders the product) and the developing project (who implements it) [A21]. Thus the architectural documents contain requirements-related information.

    Customer input is sought for complex functionality, through interviews and direct communication between stakeholders and senior engineers [A21]. The high requirements level provides the development engineers with freedom to design innovative solutions (EB4), which the system architect believes has led to producing very creative products [A21]. However communicating these details to system testing is challenging (EC4) [A21] and lack of requirements information forces testers to elicit further requirements, primarily from working software [A21]. The testers' requirements understanding of functional and, in particular quality requirements, is primarily obtained by executing the source code [A21] and then documented as test cases [A21]. Thus, a lack of requirements documentation during implementation is compensated by developer-tester interactions, in particular when the test case requirements fail [A21].

Similarly, requirements are validated through close interaction between roles (EB1), in particular between the product manager, systems architects, developers and testers [A21]. For complex functionality, customers are provided with working software for which they provide feedback and validation [A21].

### 5.1.2　　Verification: Company A

The product requirements specification [A21] and the architectural documents [A21] provide input to the testers in designing test cases. However, due to incompleteness, inaccuracy and unavailability of this requirements information, additional work is required to identify the full and correct requirements necessary to verifying the product (VC2) [A21]. In addition, documenting TCRs as part of the testing process makes verification uncertain since the requirements are less formal and more vulnerable to being changed as part of updating test cases [A21]. Manual system tests [A21] and automated function and unit tests [A21] are used to verify the implementation. The function test cases are used by the system architect in selecting the test scope when integrating new functionality (VB1) [A21]. The company wants to increase the use of automatic testing since this is more efficient [A21] and less prone to variability in testing due to who performs them (VC1) [A21].

When implementing a change, test cases for the affected software areas provide an executable specification for verifying that the legacy functionality is intact [A21]. The TCR practice's support for regression testing (VB1) supports efficient regression testing of the many products based on the company's software product line [A21].

### 5.1.3　　Tracing and Managing Changes: Company A

Requirements changes can be proposed by any project role through a change requests [A21], while issue reports are used for smaller changes [A21]. Test cases are used by the product manager to analyse the impact of issue reports and change requests (MB4) [A21]. Even so, much time and effort is spent on discussing and identifying the actual requirements; 'how it is supposed to work' [A21], since there are no documented requirements for around half of the incoming issue reports [A21]. Thus, there is insufficient coverage of the full set of requirements expected by the customer and the users. Requirements coverage is considered during the initial test design, but not kept updated due to weak tracing between TCRs and the formal product requirements specification [A1, A2, A3]. In addition, the test cases lack requirements-related information (MC2) needed when a test case fails [A2] or is updated [A3], e.g. priority and stakeholders. The challenge of insufficient requirements coverage is partly mitigated by adding test cases for resolved issue reports [A21]. As a product manager said: 'That's how we record the new requirement.' [A21] However, as the test lead said 'we cannot test everything' [A21].

Communication of requirements changes is a challenge at Company A where requirements are not documented as test case upfront, but rather after they have been implemented [A1, A2]. In particular, the system testers suffer from not being aware of all changes (MB1) [A21]. They can then not add or update the corresponding test cases (MB2), which then causes (incorrect) test cases to fail for intended behaviour [A21]. In particular, this is the case for quality requirements which are difficult to specify and communicate on paper [A21].

The product line set-up used to develop many products on the same software base causes challenges in managing requirements variations between the products both in general, and specifically using TCR (MC3). Since the tool infrastructure does not support managing such variations in the test cases (TC1) the requirements variations are currently handled by noting these directly in the test cases, information which everyone does not read [A21].

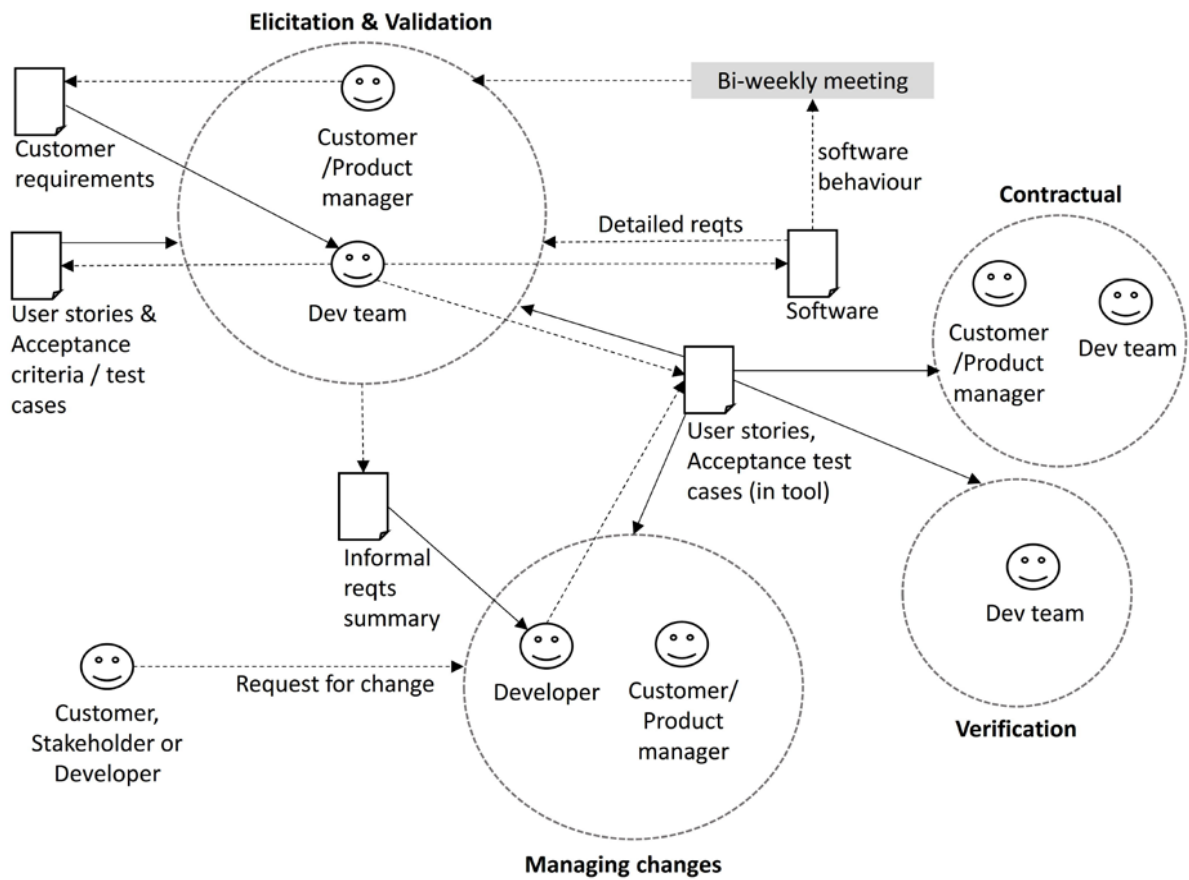

**Figure 2.** Overview of requirements flow for Company B's product development using tool-supported behavior-driven development. Notation described in Section 4.2.4.

### 5.2 Company B: Tool-Supported Behaviour-Driven Development TCR

Company B actively applies the TCR practice through behaviour-driven development supported by tools such as Cucumber and FitNesse. The customer and the product manager (often the same person) define product and customer requirements, see Figure 2. Based on these the development team identify user scenarios or stories that describe 'how this feature can interact with a customer or another system' [B2]. Then, for each iteration, the development engineers produce acceptance criteria and acceptance test cases from these requirements [B1, B3]. These criteria describe scenarios using a domain-specific language [B2]. There is a close working relationship between the product owner and the development team. The product owner is available to the development team who is usually located at the customer's site.

#### 5.2.1 Elicitation and Validation: Company B

Requirements are elicited through directly communication between business and engineering roles around acceptance test cases (EB1) [B1]. Requirements validity is ensured by bi-weekly meetings [B1] through which the customer and the engineering roles 'understand each other even though we are not on the same technical level' (EB2) [B3]. A working version of the software was mentioned as further supporting this interaction [B3]. The communication is also facilitated by the engineers adapting higher and less technical level of communication [B3] and adapting to domain terminology in the code design (for the test cases) [B2]. Furthermore, the specific format used for requirements raised the technical discussion to a more conceptual level; to communicating more about needs and goals than about solutions (EB3) [B3].

This communication level was seen to enable non-technical experts to participate in the elicitation, e.g. usability designers [B3]. However, customers find it hard to be fluent in the structured format for acceptance criteria (EC3) [B3], similar to a domain-specific language.

The elicited requirements are primarily documented as acceptance test cases in a tool. Interviewee B1 stated that the clarification of requirements and specification of test cases 'go well together because we understand more of the requirement. They concretized what we will do' [B1]. In complex cases (EC4), informal summaries are also produced [B1]. The interviewees described that acceptance test cases are useful as a system specification. However, interviewee B3 indicated that they are not fully stand-alone, i.e. the documentation can be read 'to get an impression. But, if you wonder what it means, you look at the implementation.' [B3]

Over time the company has achieved active customer involvement and effective requirements communication with this practice (EB1, EB2), despite challenges to ensure this involvement and that 'we [customer and development team] spoke the same language' (EC1) [B3]. Interviewee B1 said that active business involvement strengthens requirements validation and ensures a common view of requirements (EB2) [B1].

Similarly, it is a challenge to get business roles to write or revise the user stories or the acceptance test cases (EC2), in particularly directly in the tools [B3]. Interviewee B1 believes this requires technical skills (EC3) [B1]. This may also be due to the customer not trusting the technical systems or seeing the value in working on the same requirements base [B3]. However, the added benefit of customer directly editing requirements is believed to be smaller than the cost of training them to do this [B1]. The company has also experienced issues with setting up common access across networks [B1].

The interviewees see that customer competence affects the communication and the requirements picture produced with this practice (EC3) [B3]. In particular, non-technical customers seldom focus on quality requirements (EC4) [B3]. In contrast, engineers find it hard to match high-level requirements to the code required for automatic test cases [B1].

Complex interactions and dependencies between requirements (EC4), e.g. for user interfaces [B1] and quality requirements [B2], are a challenge to capture with acceptance test cases. For user interfaces the issue is connected to the complexity of between components. All interviewees mentioned the challenge to motivate engineers to write acceptance-level test cases [B2].

### 5.2.2 Verification: Company B

One of the main benefits of TCR stated by the interviewees is the strengthened alignment of business requirements with verification of the agreed requirements (VB2) [B1] and support for efficient regression testing (VB1) [B1]. The acceptance test cases verify that the system meets the requirements that were agreed as acceptance criteria [B1]. Similarly these test cases (connected to acceptance criteria, i.e. requirements) allow for status tracking based on requirements coverage (VB3) [B2].

The executable specification provided by this practice, in combination with unit tests, acts as a safety net that enables projects to 'rebound from anything' [B1] by its support for regression testing (VB1) which allows for efficiently managing bugs and performance issues. This makes engineers confident in frequently releasing production code; weekly for the described project. Interviewee B3 said that with this approach projects 'deliver on time and almost on budget' [B3]. The set-up has enabled projects to efficiently manage bugs and performance issues.

However, there are also challenges in creating automated tests and achieving a good executable specification [B1]. The right balance between acceptance and unit test cases needs to be found. Interviewee B3 described that 'writing too many acceptance criteria that deal with smaller things' can be very costly if they need changing later on [B3]. Furthermore, for quality requirements there is a challenge in automatically testing performance and other quality aspects on actual hardware and in a live testing environment (VC3) [B1].

### 5.2.3 Tracing and Managing Changes: Company B

The alignment of business and technical aspects through TCR is supported when managing requirements changes by the use of acceptance test cases as formal requirements (MB3) [B2, B3], thus having implicit requirements-test traces (TB1). Any agreement on changes made, e.g. in face to face meetings, must be reflected in the test cases before they become an actual change (MB2) and at the end of a project the acceptance test cases show 'what we've done' [B2].

The support for regression testing provided by TCR (see Section 5.2.2) 'limits the danger of changing something' [B2] and thus acts as a risk mitigator when managing requirements changes since the impact of changes are automatically caught (MB4) [B3].

Locating requirements information in TCRs, e.g. when performing impact analysis for a requirements change, can be hard for badly structured acceptance test cases (MC1). This requires tacit knowledge [B3] without which it is time consuming to locate the requirements. Interviewee B3 suggested that the tools could be extended to support searching for certain logic rather than just syntax.

Maintenance of the acceptance test cases is an issue that needs to be considered when applying this practice [B1, B2, B3]. Interviewee B3 pointed out that test cases are more rigid than requirements and thus more sensitive to change [B3]. The same issue surfaces when new requirements affect old ones or when requirements misunderstandings are detected resulting in updates to the (old) acceptance test cases. Thus, there is a risk of deteriorating test case quality and subsequently also to requirements quality when testers make frequent fixes to get the tests to pass [B2].

Interviewee B1 described that when an acceptance test case fails an engineer analyses and discusses this with a business role, ideally the one originally involved [B1]. However when larger changes are needed interviewee B3 had experienced that this was best handled by the developer themselves using requirements documentation in the form of the existing test cases and the informal requirements summaries to guide them in revising the requirements aspects of the acceptance criteria (MB3) [B3].

### 5.2.4 Customer Agreement: Company B

The agreed requirements documented as test cases are seen as a contract between the customer and the development project. If the customer later proposes conflicting requirements, the originally agreed acceptance test cases have then been referred (CB1), thus supporting the project in the customer dialog [B3].

## 5.3 Company C: Failed Story-Test Driven TCR and Stand-Alone Manual TCR

The TCR practice was introduced as part of the agile transition that was ongoing in 2009/10, however our reinvestigation shows that the practice is limited to the self-certification of customer requirements and for internal API development (see Section 5.4). For software development in general, the only aspect of TCR found to be applied is as support for regression testing, where test cases are viewed as requirements documentation for legacy functionality. Thus, test cases are not used as requirements during development of new functionality. The initial intention was to define requirements as user stories and acceptance test cases within a team consisting of a product manager, developers and testers. Acceptance criteria were to specify 'how the code should work' [C8] and be documented by testers as acceptance test cases traced to user stories. Another team was to maintain the software including the user stories, test cases and traces. Integrating the differing characteristics and competences of RE and testing was seen as a major challenge [C5, C10]. Tool support for integrating RE aspects in the test cases was needed for noting requirement source, connections, dependencies, and validity for different products [C5].

The current requirements flow (shown in Figure 3) partly corresponds to the one planned, although user stories and test cases are no longer defined upfront and the planned tool support for connecting these artefacts has not been put in place (TC1) [C21] and was mentioned as one reason for not fully implementing TCR. Despite being a large company, requirements communication is primarily managed through informal communication with the development teams although complemented by documentation, see Figure 3. Documentation is primarily used internally for scoping of high-level requirements, and towards customers to agree on requirements for which customer requirements specifications and requirements compliance documents are used. In addition, some customers provide a certification test suite that is to verify support for the customer's requirements. Internally, feature definitions and user interface specifications are used as requirements documentation by the development teams. For large and complex customer-specific requirements, the development teams have direct communication with the customers to clarify and detail requirements. This interaction often also involves providing the customer with working software.

During implementation high-level requirements are detailed by the development team in collaboration with the product owner and a usability designer. These detailed requirements are implemented and function and unit tested by the developers. Once the software is integrated into the software product line it is system tested and the customer certification test suite is executed.

The main reason for moving away from the TCR practice, mentioned by a senior systems architect, is a shift in priority from quality to speed [C21]. Faster software update channels enable the company to 'act fast, fix them [the problems] quickly' [C21] and frequently release new and improved software to the market. In addition, the development teams are freer now to decide for themselves how they work [C21] with little documentation mandated by the development process [C21], which has resulted in the current process.

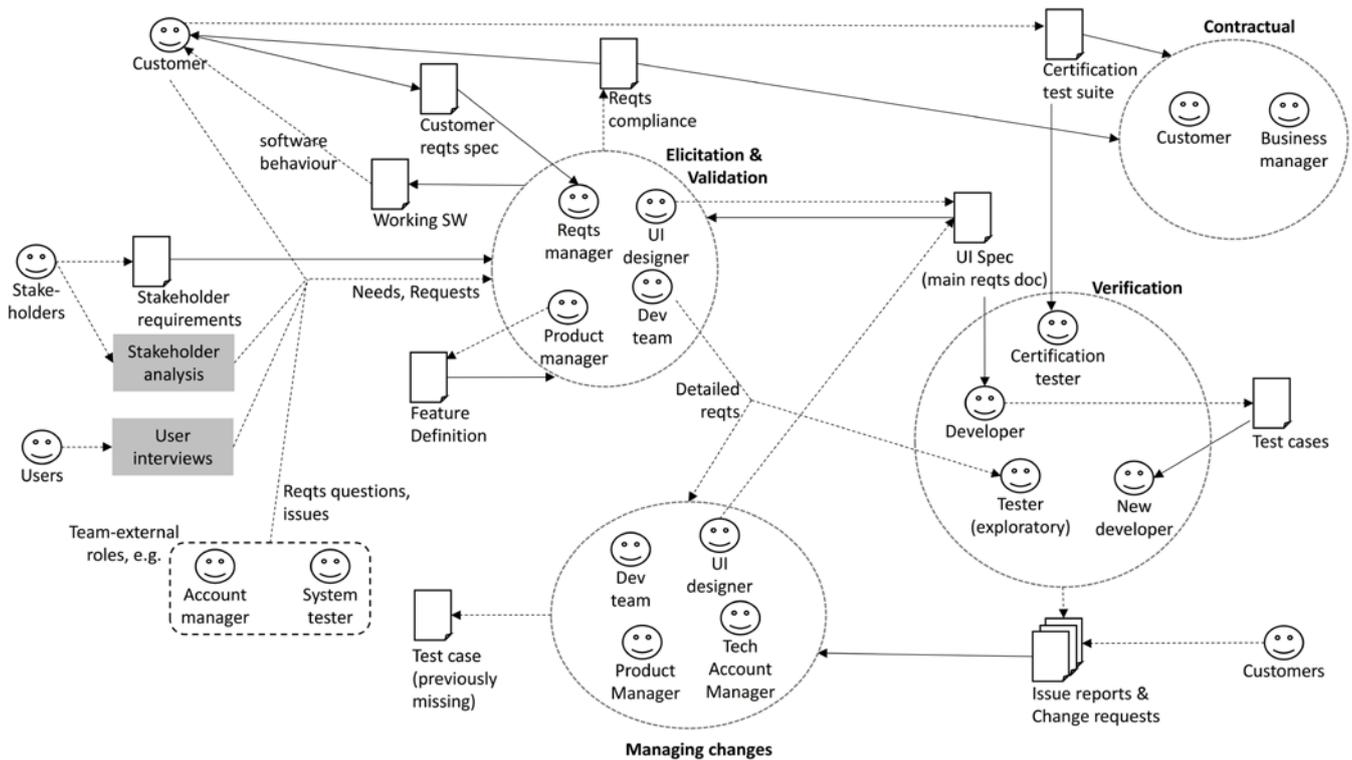

**Figure 3.** Overview of requirements flow for the software product development within Company C's. The notation is described in Section 4.2.4.

### 5.3.1 Elicitation and Validation: Company C

Development teams elicit and validate requirements in close collaboration with a product manager (EB1) [C21]. The formal input consists of a one-page document (feature definition) describing the business case and top-level scenarios, and for customer functionality there is a high-level customer requirements specification [C23]. This high level of formal requirements input (often at vision level) [C21] provides the development teams with freedom in detailing requirements and encourages them to contribute with their creativity (EB4). Further input is elicited through stakeholder analysis [C21], interviews with users [C21] and direct interaction with customers. Additional requirements are identified throughout the development life-cycle in an exploratory fashion [C21], and 'discovered when implementing' [C21] rather than defined upfront before design [C21]. The product manager role is vital in facilitating the exploratory requirements elicitation, but the role requires a combination of technical and business knowledge which is hard to fill (EC3) [C21]. The requirements are validated by customers and other stakeholders by using working software [C21]. Several participants stressed that it is expensive [C21] and 'hard to write any test cases at all beforehand' [C21]. This would require the stakeholders to be able to express their requirements as test cases (EC3) [C21].

   The development team produce a user interface (UI) specification [C21], and unit test cases in-line with the implemented requirements. Defining test cases for user interactions was mentioned as a challenge (EC4) [C21]. Rather, the UI specification is seen by the development team as 'our requirements document … that lives together with the development.' [C21] Albeit this document is not used to communicate requirements outside of the development team, although it was suggested to 'be a good start for communicating.' [C21] This lack of requirements documentation leads to a large amount of questions from team-external roles concerning how the software is supposed to work, i.e. the requirements [C21], more questions than they can answer [C21]. However, the participants do not see this as a problem but rather an opportunity to elicit future requirements based on this feedback [C21].

   Concerning communicating requirements through TCR some participants believed they could provide 'some help to understand' detailed requirements [C21]. Although, as the technical account manager said 'a test case is not as clear to read as a requirement' and it 'takes longer to understand a test case than to read a requirement' [C23]. Furthermore, one systems architect pointed out that a dedicated tester with less coding experience may 'have a hard time to "read" the functionality from the [unit test] code' (EC3) [C21]. Thus, (automated) test cases require a certain level of technical competence to read and understand as requirements. Similarly, defining TCRs was believed to require specific test competence, not generally held by developers [C21]. This competence includes defining the TCRs as requirements, rather than 'how you implement it [C23].

### 5.3.2    Verification: Company C

The developers perform function testing and automated unit testing [C21] once 'the right requirements' [C21] have been identified (VC2). Since these requirements are not documented they can only be verified by the development team [C21]. System-level testing, e.g. of performance, is performed by team external testers, often located at a different site [C21]. Exploratory test missions are used for function testing, i.e. step-by-step instructions for high-level use cases [C21]. These test missions are less time consuming to write and maintain than automated test cases at the use case level [C21]. However, test missions require the same skilled people to execute them to get reliable (and repeatable) results (VC1) [C21]. The combined set of test cases acts as an executable specification that facilitates re-testing after refactoring [C21] and is considered the main benefit of TCR (VB1) [C21].

Finally, the customer test suite is used for self-certification of the customer requirements [C23] with the added intention of catching issue reports and change requests from the customer early on, thereby increasing development efficiency [C23]. However, this has not been achieved. While the cost of executing these (primarily) manual test cases is very high, this testing uncovers very few issues [C23]. Even so, the customer later reports very many issues during their exploratory customer acceptance testing [C23]. The problem, according to the interviewee, is the weak quality of the self-certification test suite (VC2) mainly caused by low requirements coverage [C23], in combination with a low rate of change [C23], i.e. not kept updated. The test suite does not have good coverage either of the formal requirements or of the informal (previously uncommunicated) requirements represented by the many issue reports filed by the customer [C23]. In addition, these reports contain issues on non-agreed requirements [C23], and are thus actually change requests.

### 5.3.3    Tracing and Managing Changes: Company C

Requirements changes are triggered by change requests and issues reports [C21]. The analysis of changes are handled mainly through communication within the development team [C21]. The developers are often involved also in customer issues even though these are analysed by a technical account manager, since 'only the developers know what they have developed.' [C23] The (undocumented) customer requirements received through issue reports are documented by the company by adding the corresponding test cases (MB2) [C23]. The test cases are used to catch the impact of implemented requirements changes (MB4) [C21] rather than as documentation that supports impact analysis [C21]; 'It [the impact] will be detected when you run them [the test cases].' [C21]

The lack of communication of implemented changes to roles outside of the development team was mentioned as a problem, in particular for system testers at remote sites and customer communication. A system architect said: 'they [team-external roles] don't know if it is a bug or a feature change….Well-documented requirements would solve that.' [C21] However, using TCR for communicating requirements changes is seen as too costly and error prone [C21] and, from the development team perspective, the value of maintaining TCRs is perceived as low [C21].

Traces exist between the customer's test cases and the corresponding formal requirements, although tracing is not complete. Furthermore, different versions of the customer requirements specification often cause issues when the customer acceptance test is performed based on a newer version of the specification (MC3) [C23].

Furthermore, the company's product line software runs on several different hardware versions and poses a challenge in managing requirements changes as test cases (MC3) [C21]. A test case may fail due to a variability in the underlying hardware rather than that the requirement is not fulfilled. In this case you might 'fix the test case and at the same time you might rewrite the requirements' [C21]. Thus, applying TCR may cause the requirements to become less formal and more prone to inadvertent scope creep.

### 5.3.4    Customer Agreement: Company C

The certification test suite is an example of TCR used for contractual purposes [C23]. In order to deliver the final product the company must show that this test suite has passed (CB2). In addition, the agreement is documented through the compliance statement provided by the company. At the detailed level informal agreements are made based on direct communication between customer and the development team around working software and user interaction flows [C23].

## 5.4 Company C: API Development with Stand-Alone Strict TCR

Company C applies a TCR process for developing APIs (application programming interfaces) for their large code base. When a development team needs a new API to an existing component, the team's software architect request that a new API is specified by a systems architect [C22]. This new API is designed and specified as executable test cases (TCR); an API specification test suite. This test suite is then used for developing software that invokes the API and when implementing the API itself, often by different development projects and involving different geographical sites. The API specification test suite ensures that the API implementation and the software that uses it match [C22], thus that the API requirements are met.

### 5.4.1 Elicitation and Validation: Company C API Development

The systems architect elicits and validates the needs and requirements on a new API through direct communication with the requesting software architect (EB1) [C22]. The API architect then specifies the API requirements as automatic test cases (API specification test suite) [C22]. This test suite includes documentation of parameters and method prerequisites [C22], 'to explain how the API is intended to work.' [C22] This takes around two weeks. [C22] The API specification test suite is informally reviewed by the requesting development team, and formally approved by a governance board of system architects, i.e. the API board. This review process takes two weeks at the most. [C22]

The API specification test suites are used to transfer requirements knowledge. The test suites are read by developers and their source code reused to initiate developing. The test suite enables them to 'see how it [the API] is intended to work' [C22].

The TCR practice works well in this context where all involved roles have similar technical competence (EC3), thus reducing the effort required. As the interviewee said: 'those involved can describe their world with source code.' [C22] In addition, the API specification test suite supports communication and coordination between several project teams [C22]. Thus, upfront specification of requirements as test cases is worthwhile for this context even though it is a slower, and more costly and bureaucratic way of working [C22]. The interviewed systems architect suggested extending the TCR approach to the product level as supporting a high-level specification structured by use cases. [C22]

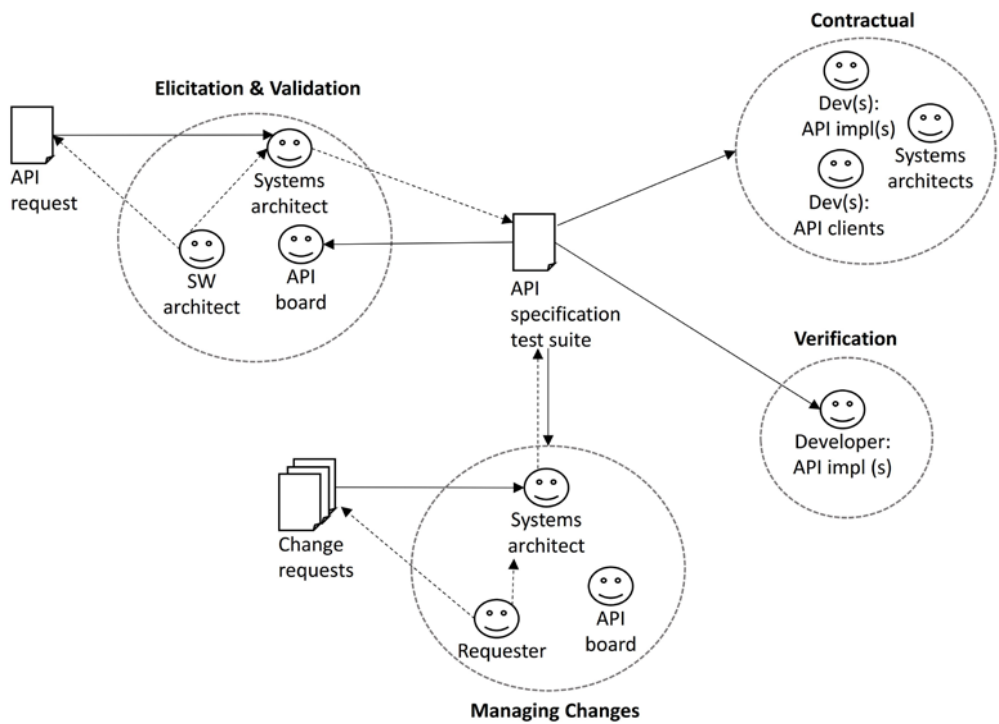

**Figure 4.** Overview of requirements flow for Company C's API development. The notation is described in Section 4.2.4.

### 5.4.2 Verification: Company C API Development

The API specification test suite is executed by the developers implementing the API, and thus used to verify compliance of the API implementation. [C22] In addition, the test suite is used for regression testing each new delivery. [C22]

### 5.4.3 Tracing and Managing Changes: Company C API Development

Changes to an API specification test suite, including bugs and issues found during development, are handled in a similar way to requests for a new API. Thus, all changes are approved by the API governance board and communicated through updating the API specification including its test cases (MB1, MB2) resulting in maintaining alignment between the API requirements and test cases (MB3). New releases of an API are version controlled and for larger (non-backwards compatible) changes an API can be deprecated and replaced by a new one [C22].

For changes, the API specification test suite is used for regression testing, i.e. for after-the-fact impact identification (MB4) rather than upfront impact analysis [C22]. When problems are found the traces to the source code (TB1) are used to analyse the issue [C22].

The interviewed system architect pointed out that there is very limited change and maintenance needed for the API test suites [C22], which is one reason why the approach is successful in this context.

### 5.4.4 Contractual Agreement: Company C API Development

The API specification test suite is the contract between the system architects and the developers involved in implementing the API and in using it, i.e. implementing the clients of the API [C22]. For the developers using the API, the API specification test suite conveys requirements regarding the API and how to invoke it. The API specification test suite also ensures consistency over multiple implementations of one API (CB2), e.g. for multiple software configurations each implemented by different developers [C22]. The API specification test suite supports independent and parallel development [C22] and reduces the risk of inconsistencies and varying interpretations of the requirements (CB1).

## 6 Discussion: TCR Supporting RE and TCR Variants

We will now answer our two research questions by consolidating and discussing the results from the three case companies. The limitations of these findings will also be discussed.

### 6.1 How Test Cases of the TCR Practice Can Support the Main Requirements Activities (RQ1)

For the TCR practice, test case artefacts are central in the coordination between the different roles involved in software production. This is illustrated in the depicted requirements flows of the studied companies, see Figures 1-4. We will now discuss how the test cases of the TCR practice can support the main activities of RE, i.e. how the test cases fulfil the role that the requirements specification traditionally fills as defined by Lauesen (2002), namely in supporting the elicitation and validation of stakeholders' requirements, software verification, managing requirements including tracing, and as documentation of customer agreement. In answering the question of 'how?' the practice is applied similar benefits and challenges were found for our case companies. These are summarised in Table 7 and described below in the context of how TCR support requirements activities. Furthermore, the context and specific expression of these can be located in the (previous) results section using the references inserted therein to individual benefits and challenges.

**Table 7.** Summary of benefits and challenges of using test cases as requirements for the main requirements activities (based on Lauesen 2002). Coded by RE activity (E, V, T, M or C), (B)enefit or (C)hallenge, and sequence number.

| Benefits | Challenges |
|---|---|
| **Elicitation and Validation** | |
| EB1 Cross-functional communication | EC1 Good Customer-Developer relationship |
| EB2 Align goals & perspectives between roles | EC2 Active customer involvement |
| EB3 Address barrier of specifying solutions | EC3 Sufficient technical and RE competence |
| EB4 Creativity supported by high-level of requirements | EC4 Complex requirements, e.g. quality requirements |
| **Verification** | |
| VB1 Supports regression testing | VC1 Varying (biased) results for manual tests |
| VB2 Increased requirements quality | VC2 Ensuring correct requirements info to test |
| VB3 Test coverage / RET alignment | VC3 Quality requirements |
| **Tracing** | |
| TB1 Implicit Requirements - test case tracing | TC1 Tool integration |
| **Managing Changes** | |
| MB1 Communication of changes | MC1 Locating impacted requirements |
| MB2 Requirement are kept updated | MC2 Missing requirement context |
| MB3 Maintaining RET alignment | MC3 Multiple products in one product line |
| MB4 Detecting impact of changes | |
| **Customer Agreement / Contractual** | |
| CB1 Facilitate resolving conflicting views | CC1 Use-case related structuring |
| CB2 Support certification of compliance | |

### 6.1.1 Elicitation and Validation

Requirements documentation in the form of test cases is mainly used for elicitation and validation at Company B and for the internal API development process at Company C, while test case are only partly used for this purpose at Company A, namely for quality requirements. However, the practice also includes the benefit of *direct and frequent communication between business and technical roles* (EB1) which further supports requirements elicitation and validation. This was observed for all of our case companies. This frequent direct interaction supports the integrated RE approach strongly connected to agile development principles (Sommerville 2005). The customer involvement in combination with (for Company B) agreeing to requirements at the acceptance test level provides the benefit of *aligning the customers' goals and perspectives with those of the technical roles* (EB2) and thus supports defining valid requirements. This confirms observations by Melnik and Maurer (2007), Park and Maurer (2009), Haugset and Hanssen (2008) and Latorre (2014). Furthermore, at Company B, the use of the acceptance criteria format led to customers expressing requirements at a higher abstraction level instead of focusing on technical details. Thus, expressing requirements as test cases has the benefit of addressing the *elicitation barrier of requesting specific solutions* (EB3) rather than expressing needs (Lauesen 2002).

At Company A and Company C, involving the technical roles in the detailing of high-level requirements *induces creativity* (EB4) in designing innovative and cost-efficient product features. This confirms findings by Bjarnason and Sharp (2015b) that a difference (distance) in abstraction level between requested requirements and the final software behaviour encourages development creativity. For market-driven companies like Company A and Company C this provides an important competitive advantage that enables them to differentiate on the market.

Nevertheless, the TCR practice requires *good customer relations* (EC1) and *active customer involvement* (EC2), which Company B has achieved but only after investing time and effort on this. Even so the development team, rather than the customer is still the ones directly managing the detailed requirements, i.e. the acceptance test cases. This is believed to be due to *insufficient technical competence* (EC3) to manage requirements as this level of technical detail. A contrasting example is Company C where test cases are an efficient way of communicating requirements since only technical roles are involved. *Active customer involvement* is a known challenge for agile RE due to time and space restrictions for the customer, but also due to that the customer or customer-proxy role requires a combination of business and technical skills (Ramesh 2010, Kongsli 2006). In particular, using TCR to define requirements before implementation (as opposed during the testing process) is facilitated when *those involved all have technical roles* (EC3).

For this reason the practice works well when only technical roles are involved as for API development at Company C, while TCR is not applied for their product development where the product managers do not 'speak source code'.

Using TCR for *complex requirements* (EC4) such as usability and quality requirements is a challenge found for all three companies. This is due to the complexity of these requirements and to the need for the additional competence of the roles involved. All three companies described the challenges of documenting complex interaction, e.g. between components, as test cases. For Company A and C, the cost and difficulty of expressing, e.g. usability requirements, as test cases was mentioned as a reason for not documenting requirements upfront (before implementation) as test cases. Company B also mentioned that limited technical knowledge affects the customer's ability to discuss *quality requirements*. This can lead to neglecting to elicit them altogether (Ramesh 2010). Similarly, capturing complex requirements with acceptance test cases is a challenge, mentioned for user interactions and quality requirements. However, the close relationship between requirements and test cases is utilized by Company A and Company C to facilitate eliciting quality requirements from working software.

Business domain tools can be used to facilitate the customers in specifying acceptance tests (Park 2009). For example, Haugset and Hanssen (2008) report that customers used spread-sheets to communicate information and never interacted directly with actual test cases.

### 6.1.2 Verification

The TCR practice supports verification of requirements by supporting *regression testing* (VB1), as seen for all three case companies even for those using manual test cases. For example, Company A experiences these benefits in ensuring the agreed requirements which are documented primarily as manual test cases. However, manual test cases are more expensive to execute, as for the self-certification example for Company C, and also more vulnerable to *variations in execution results* (VC1) depending on who performs the testing. This latter aspect is a challenge at both case companies that use manual test cases, i.e. Company A and C. In contrast, automatic test cases are more efficient and Company B uses automatic acceptance tests as a safety net that catches problems and enables frequent release of product-quality code. This was also observed by Kongsli (2006), Haugset and Hanssen (2008), and Latorre (2014). The support for regression test is especially valuable for ensuring legacy and product-line functionality. However, there are also challenges in efficiently verifying multiple parallel products based on the same product-line code base. Tool support for managing the requirements variabilities when applying TCR is missing for Company A and Company C, which both have a large product-line software base.

The TCR practice supports *requirements test coverage* (VB3) when all requirements are specified as test cases and then verified, as seen for Company B. In this case, when the practice is applied with tool support the requirements coverage can also be measured by executing the test cases.

*Requirements quality* (VB2) is supported by the TCR practice since test cases are per definition verifiable. In addition, the format used by Company B supports defining clear requirements and defining the TCRs in connection with the elicitation (as for Company B) ensures that the agreed requirements details are captured and validated. This is an important aspect for the verification effort, which relies on correct, verifiable, clear and unambiguous requirements (Davis 2005).

In contrast, documenting requirements details later (after-the-fact of implementation), as part of the testing process, poses communication issues at Company A and Company C due to *incorrect requirements information* (VC2). For these companies it is a challenge to ensure that the requirements knowledge elicited by the development team is communicated to the testers. Communication gaps result in test cases that suffer from similar quality issues as are known for traditional requirements specifications (Bjarnason 2011a), e.g. incompleteness and outdated requirements.

This is experienced by Company C concerning the customer test suites used for self-certification. In which case the incompleteness of these test suites results in failure to capture the majority of the customer issues in the certification testing. However, test design and selecting a cost-effective test level also cause challenges in achieving good requirements quality through test cases. For Company C the developers who perform the unit and function testing lack specific test competence, which tends to result in more white-box than black-box tests, thus more solution-oriented than requirements-oriented test cases. In addition, expressing TCRs at acceptance or unit test level affects the cost of creating and maintaining these test cases, something that had been considered for Company B and for Company C.

Companies B mentioned *quality requirements* (VC3) as a particular challenge in automatically testing these for embedded devices as this requires actual hardware. This confirms previous findings by Ramesh (2010) and Haugset and Hansen (2008), i.e. that quality requirements are difficult to capture with automated acceptance tests.

### 6.1.3 Tracing

Tracing of requirements and test cases is supported by the TCR practice, however the benefits depend on the context. Merely using test cases as de facto requirements at the detailed level, as in Company A, does not affect tracing since certain requirements information, e.g. priority and stakeholder, is not noted in the test cases but in a (separate) requirements specification. Furthermore, when the TCRs to a large extent replace the formal requirements specification as the actual and most updated source of requirements, as for Company A, it is not always possible to trace to the un-updated (traditional) requirements specification.

For the behaviour-driven development approach applied at Company B, the tools *implicitly trace* (TB1) acceptance criteria and test cases, although there are no traces between the original customer requirements and the acceptance criteria. Hence, as the requirements evolve this knowledge is reflected purely in the test cases (Mugridge 2008).

### 6.1.4 Managing Changes

The TCR practice, when consistently applied, provides benefits in managing requirements in an efficient way throughout the development life-cycle. As mentioned for Companies A and B, the practice facilitates a joint understanding of requirements that provides a base for discussing and making decisions regarding changes. The practice, thus supports *communication of changes* (MB1) although this requires effort in involving development engineers in the requirements discussion. When requirements are documented upfront (before implementation) this benefit is even more prominent, as seen in Company B and for API development at Company C. In these cases, the benefits also include *requirements being kept updated* (MB2) and *aligned with test cases* (MB3) through the integration of requirements and test cases. This is also seen to some degree for Company A where the test cases act as the requirements and are kept updated. Keeping requirements updated after changes is a known challenge (Bjarnason 2014). In contrast, when the test cases are not kept aligned with the formal requirements this causes additional work with managing and resolving issue reports on correctly implemented functionality. This was seen for the customer's self-certification test suite used by Company C and at times at Company A when the developers did not communicate changes to the testers.

When the requirements are documented in an executable format, conflicting new or changed requirements are likely to cause existing test cases to fail, thus test cases can be used to *detect the impact of changes* (MB4). All three companies utilise this benefit by catching the impact after having implemented a change rather than through upfront change impact analysis, as part of the change decision process.

*Locating requirements in a set of test cases* (MC1) was mentioned as a challenge for Company B due to badly structured test cases. The difficulty of organizing and sorting automated tests has also been reported by Park (2010) and Erman (2015).

*Contextual requirements information* (MC2), e.g. purpose and priority (Lauesen 2002), is seldom retained in the test cases. This was described as a challenge for Company A in managing requirements changes. This kind of information can support impact analysis and managing failed test cases. Without access to contextual information from the test cases, additional effort is required to enable informed decision making.

For all three case companies we see that *suitable tool support* (TC1) is required for the practice to work well. At Company B this is achieve, while this is not the case for Company A and Company C. At Company C challenges in implementing *tool support* for connecting requirements and testing is one reason why TCR has not been implemented as planned. As part of the agile transition the intention was to detail user stories into acceptance test cases and to retain traces between them as a manual, albeit straight forward task. The responsibility for these traces was clearly defined in the development process, a practice identified by Uusitalo (2008) as supporting traceability.

Using the TCR practice for developing many different product variant based on a *common product line* (MC3) requires adequate TCR tool support. This is experienced as a challenge at Company A, where requirements variations are managed by noting these directly in the test cases, which later poses problems since not everyone reads this information. Company C also mentioned this challenge related to managing requirements variations for multiple products in a product line using test cases.

### 6.1.5 Documentation of Customer Agreement

For all three companies, the TCR practice was described as providing documentation of the agreement between the customer and the developers during elicitation. This documentation can be beneficial in later development stages, e.g. to *resolve conflicts regarding requirements* (CB1) (e.g. scope or interpretation). An example of this was found for the behaviour-driven development approach applied by Company B. In this case the specific and clear format for defining these TCRs was described by the engineers as being very useful documentation to refer to when the customer requested different (non-agreed) requirements, which could then be resolved by consulting the TCRs. Similarly, for API development at Company C the TCRs provided a unified interpretation of the requirements ensuring consistent understanding and implementation of them throughout the company.

Another situation where test cases can act as a formal agreement is in *certifying that an implementation is compliant* (CB2) to a standard specification, as is the case with customer requirements for Company C. In this case, the customer uses the same certification test suite for multiple customers to manage and ensure similar functionality between manufacturers. The TCRs fill a similar purpose for the API development at Company C by requiring that these test cases pass before a new software version is accepted internally.

TCR documentation is a suitable format for formal requirements agreements, in particular when interacting with a technical customer. This is the case for the API development within Company C where the 'customer' and the 'requirements engineer' are both software architects who can communicate via source code. In this case the TCR practice was well established while it had failed to be implemented for software development in general at this company.

The firm and structured format used for Company B as part of behaviour-driven development seems to provide a clear way of documenting requirements. In contrast, it was suggested that the readability of Company A's de facto TCR documentation was a challenge since the *requirements information per use case* (CC1) could not be easily extracted. With a use-case related structure of the TCR documentation the accessibility of these requirements may be improve and their use extended to customers and other non-technical roles.

### 6.2  Variations in Applying the TCR Practice (RQ2)

Each case company uses the TCR practice differently, leading to variations in how the practice is applied. Our description of these variations is structured around four facets of the practice that emerged while performing a cross-case analysis of the results. These facets facilitate a comparative description of the practice variations across different applications, and are as follows:

1. the *documentation time frame* for defining the TCRs, either upfront during elicitation and validation, or after-the-fact during the testing process
2. the *requirements format* used for the TCRs, which ranges from a formal domain-specific language structure to natural language test cases
3. the *machine executable specification* aspect of the practice that allows the requirements to be verified automatically
4. the *tool support* used to facilitate the main requirements activities when applying the practice

The application of the TCR practice varies over these facets for the studied cases. The *time frame* within which the requirements are captured as test cases varies between being documented upfront before implementation, and during the test process, i.e. after-the-fact once the requirements have been implemented. Upfront documentation is applied as part of the behaviour-driven development approach used by Company B and in the API development process used by Company C. In addition, Company C planned to document requirements upfront as part of their transition to agile, but the practice is no longer applied there for product development. The requirements are documented when designing test cases at Company A, i.e. after-the-fact.

The *format* used to document requirements as test cases varies from a formal to natural language structure. An example of a formal structure is seen at Company B where a domain-specific language is used to specify acceptance criteria and test cases as part of their behaviour-driven development approach. At the opposite end of the scale we find the manual test cases used for product development at Company A, which require execution by the same person to ensure consistent test results. The automatic test cases complemented by natural language comments, as applied in Company C's API development process can be seen to be of an intermediate level of formalization with a combination of formal and informal requirements format.

A *machine executable specification* where the test cases (that are requirements) can be automatically executed allows for efficient regression testing and short turn-around times in verifying requirements after changes. We observe that this aspect varies from fully machine executable with automatic test cases to a combination of manual and automatic test cases. An example of fully automated is found for the TCR API specification at Company C and for the behaviour-driven approach applied at Company B. Company C is an example of a combination of manual and automatic test cases.

The final TCR facet is the *tool environment available to specifically support the TCR practice* including the combination and integration of RE and testing aspects. Among our case companies only Company B uses a tool that explicitly supports the requirements activities of the TCR practice including the requirements format and tracing between TCRs and to higher-levels of requirements. Lack of suitable TCR tool support is one of the reasons for the practice being implemented to a lower degree at Company C than what was originally planned, e.g. tracing between user stories and test cases.

Using the four facets above, we can characterize the variations of the TCR practice we have encountered as illustrated in Table 8. Based on this, we find five variants of the TCR practice namely *de facto*, *behaviour-driven*, *story-test driven*, *stand-alone strict* and *stand-alone manual* TCR. Additional variants may be envisioned by further varying the facets and by investigating additional cases.

**Table 8.** Overview of the five identified TCR variants characterized by four facets of the practice, and the case company for which the variant was observed.

| TCR Variant | Description | Facets of the TCR Practice | | | | Observed in Case Company |
| --- | --- | --- | --- | --- | --- | --- |
| | | Documentation time frame Upfront or after-the-fact | Requirements format General or structured | Machine executable specification | TCR-specific tool support | |
| de facto | A combination of manual and automatic test cases are used as the main source of requirements information, since they contain the most reliable and updated information on implemented products. | After-the-fact | General | Partly | No | A |
| behaviour-driven | Requirements are documented as automated acceptance test cases as part of the elicitation process and using a full BDD process and tool framework. | Upfront | Structured | Yes | Yes | B |
| story-test driven | Requirements are documented as user stories and acceptance criteria, for which manual and automatic test cases are later created. Tooling supports (manually) tracing between the artefacts. | Upfront | Semi-structured | Partly | Yes | C for product development, failed implementation[3] |
| stand-alone strict | Requirements for a sub-set of the functionality are specified using automated test cases with free text comments describing these. Regular tools for automated testing are used. | Upfront | Semi-structured | Yes | No | C for API development |
| stand-alone manual | Requirements are specified by a sub-set of stakeholders using manual test cases, which are then used to verify that the stakeholders' needs are met. | Upfront | General | No | No | C for customer certification |

At Company A TCR has become a *de facto* practice due to strong development and test competence, and weak RE processes. This is similar to what Wnuk (2014) reported from another case study of agile development in the embedded software domain. However, merely viewing test cases as requirements does not fully compensate for a lack of RE. Company A faces challenges in managing requirements changes and ensuring test coverage of requirements, that have spawned research into traceability recovery between source code and test cases (Unterkalmsteiner 2015). The requirements documentation does not satisfy the information needs of all stakeholders. Furthermore, staff turnover may result in loss of (undocumented) product knowledge. As size and complexity increase so does the challenge of coordinating customer needs with testing effort (Bjarnason 2014).

Company B applies the practice consciously using *behaviour-driven* TCR including tool support. This facilitates customer communication through which the engineering roles gain requirements insight. The automated acceptance tests provide a feedback system confirming the engineers' understanding of the business domain (Park 2009). However, it is a challenge to get customers to specify requirements in the automated acceptance tests tools. Letting domain experts or customers provide information via e.g. spread-sheets may facilitate this (Park 2009).

---

[3] Company C planned to implement this variant but failed. They currently do not apply TCR for their product development, but rather define semi-structured test cases after-the-fact using a combination of manual and automated test cases (partly machine-executable) and without any TCR-specific tool support.

The third variant, namely *story-test driven TCR* with tool support was seen originally in Company C where TCR was intended to be part of a transition to agile processes. Story-test driven development (Park 2010) includes collaboration around requirements between business and development roles and documenting requirements upfront as user stories and acceptance test cases. In contrast to the de facto context (at Company A), Company C envisioned this practice of *upfront* specification of requirements as test cases resulting in a *machine-executable specification* that would facilitate maintenance of the requirements throughout the development life cycle. The requirements were to be specified in a *semi-structured format* consisting of a user story template and a combination of manual and automated test cases. To achieve this, the requirements aspects of TCR needed to be supported by the test management tools including traceability to test cases. This necessary tool support was not implemented, which can partly explain why this practice variant was not implemented for product development at Company C. In addition, user interaction has a stronger focus within Company C than testing (as for Company A), which has resulted in the user interaction artefacts becoming the main requirements documentation (rather than the test cases).

The fourth and fifth practice variants are found at Company C, where *stand-alone TCR specifications are* used in a *strict* fashion to specify internal APIs and in a more *manual* fashion to enable self-certification of compliance to customer requirements (Morris 2001). In these two variants, requirements are documented primarily as test cases, i.e. stand-alone with no, or very weak, connection to a formal requirements documentation level. These TCRs are agreed upfront before implementation starts similar to traditional phase-based development. The main difference between these two related variants lies in the use of automatic or manual test cases. The internal API development is an example of *stand-alone strict* TCR where the test cases form a machine executable specification. The customer TCR requirements consisting of manual test cases which are very costly and time consuming to execute are an example of the *stand-alone manual* variant. For both of these variants, the TCR practice is used to certify that the implementation matches the requirements.

## 6.3 Limitations

We discuss limitations of our results using guidelines provided by Runeson et al. (2012).

### 6.3.1 Construct Validity

A main threat to validity lies in that parts of the analysed data stem from interviews exploring the broader area of RET (requirements engineering and test) alignment, namely the 12 interviews analysed in the first iteration. This limits the depth and extent of the results from this data to what the interviewees spontaneously shared around the TCR practice in focus in this paper. In addition, these re-used interviews do not allow for drawing conclusions based on the absence of information since specific questions about the practice were not asked. For example, concerning support for a certain requirements activity. In particular, the fact that the practice was not yet fully implemented at Company C at the time of the interviews limits the insights gained from those interviews. However, we believe that the broad approach of the original study (Bjarnason 2013) in combination with the semi-structured interviews provide valuable, though limited, insights. Furthermore, this limitation was addressed by collecting new data in the second iteration, thus complementing the data used in the first iteration. This additional data collection allowed for a focused and in-depth investigation of the practice for companies A and C. Company B was not further investigated due to lack of research access to this company, but also due to the fact that the available data for Company B in the first iteration was more extensive than for the other two companies. However, there is a risk of missing results for the behaviour-driven development variant of the TCR practices for which further in-depth studies is needed.

The varying set of roles from each case under study poses a risk of missing important perspectives, in particular, for the limited set of roles from which data was collected for Company C in the second iteration poses. The data was collected from software and systems architects, and one technical account manager. This poses a risk of missing important aspects and perspectives experienced by other roles, even if this limitation was not intended, but occurred due to limited availability of participants during the available time frame. Due to the seniority of the software architects and the close collaboration within the development teams we believe that their views align well with those of the developers. However, the system testing perspective was not investigated for Company C, but rather how TCR is used within a development team. Further investigations of how TCR is perceived by business-oriented roles, e.g. product managers, and by system testers at Company C would complement these results. In general, more research is needed into how the practice affects business versus technical roles including testers at different levels.

### 6.3.2 External Validity

The fact that our results are based on data from three case companies limits the completeness and general applicability of our findings. For example, there are most likely more variants of the TCR practice than the ones we have identified. However, we believe that the findings can be valid beyond the three investigated case companies and may be generalized to some extent to companies applying an agile development model. However, this needs to be assessed through theoretical generalization (Robson 2002, Runeson 2012) on a case-by-case basis by comparing to characteristics of the case companies (reported in Section 3) and the context of the specific implementation of the practice.

### 6.3.3 Internal Validity

As for all case studies, identifying casual connection is a major challenge. For our study, the main threat to internal validity is caused by the difficulty of identifying which part of the TCR practice or other surrounding practices that cause an observed impact. For example, when considering the support for elicitation it is hard to separate between the impact of documenting test cases as requirements and the impact of having close and frequent requirements communication. In particular, this poses the risk of drawing incorrect conclusions for RQ1 regarding how the documentation of requirements as test cases supports the requirements activities. This open risk could be addressed through more case studies and through systematic theory building based on rich(er) empirical data (Eisenhardt 2007).

### 6.3.4 Reliability

There is a risk of researcher bias in the analysis and interpretation of the data. This was partly mitigated by applying triangulation both within the group of researchers and towards the participants. For each step in the research process, the output produced by another research, i.e. transcripts, coding, summaries, and reporting, was reviewed by and agreed with another researcher. The transcripts were validated by the interviewees and participants. For the first iteration the transcripts were sent to the interviewees for validation. For the second iteration, an intermediate summary of the focus group session or interview was reviewed by the participants to correct any misunderstandings. Furthermore, a rigorous process was applied in the (original) data collection including researcher triangulation of interviewing, transcription and coding, which increases the reliability of the data selected for the first iteration. Furthermore, traces to the empirical data (interview transcripts) were retained into the reporting step by one of the authors and thus enabled validation by another author through triangulation of the reported results against the data.

There is a risk of biased results for the two additional development contexts investigated at Company C, i.e. API development and self-certification due to investigating each of these through only one interview. The risk of incomplete results for these two examples of TCR usage would be mitigated through further data collection.

## 7 Conclusions and Future Work

Coordinating and aligning frequently changing business needs is a challenge in any software development project. In agile projects this is mainly addressed through frequent and direct communication between the customer and the development team, i.e. through integrated requirements engineering (RE) (Sommerville 2005, Bjarnason 2011b). The detailed requirements are often documented as test cases rather than in a separate requirements specification, thereby reducing the effort required to keep two separate artefacts updated and aligned.

Our case study provides insights into how this practice of using test cases as requirements (TCR) supports the main requirements activities (RQ1). The results suggest that *the direct and frequent communication* enforced by this practice supports *eliciting, validating and managing* new and changing customer requirements. Furthermore, specifying requirements as acceptance test cases *allows the requirements to become a living document that supports verifying and tracing requirements* through the development life cycle. However, TCR also poses challenges in relation to requirements engineering activities, including the elicitation and verification of quality requirements, and managing changes with missing requirements context leading to difficulties in impact analysis. In all companies we have observed both benefits and challenges of TCR, explaining why in practice there is no single, commonly agreed TCR practice but variants that are adapted to the specific context and demands of the developing organization.

We have identified five TCR variants (RQ2) namely *de facto*, *behaviour-driven*, *story-test driven*, *stand-alone strict* and *stand-alone manual* TCR. For each identified variant, TCR is applied different for a set of facets of the practice. These facets can be useful in considering how to configure the practice for a specific case. The facet of *documentation time frame* influences the quality of the requirements, which in turn affects the requirements communication between project roles. In upfront TCR the documentation of test cases is done during elicitation and validation of the requirements, positively affecting requirements quality in terms of completeness through the close involvement of the business roles. Upfront TCR supports the coordination and alignment in projects with many stakeholders, for example, if certification for a standard or against customer requirements is needed. The test cases thereby serve as a firm and reliable source of requirements information, since they are used and actively maintained. *After-the-fact* TCR is a valid strategy in organizations where test cases, in contrast to requirements documents, are maintained over the product life-time and reused in different products. The facet of the *requirements format* determines the extent to which specialized knowledge is required to interpret the requirement as a customer need, as well as a mean to verify that need. Formalization reduces the risk of ambiguity and makes machine-based processing of information straightforward, however presumes a certain competence level from all involved stakeholders. For some variants, the TCR practice produces *executable specifications* that are either evaluated manually by humans or automatically by machines. This facet is important to consider in situations for which the rate of requirements changes is high, in which case regression testing with automated executable specifications can be of great value. However, the degree to which automation can be realized depends on the possibility to which the requirements documentation can be formalized, which in turn depends on stakeholder needs. Executable specifications, independently of whether they are manually or automatically evaluated, also support the certification of an implementation. Finally, certain TCR variants require that the facet of *TCR-specific tool support* is present, enabling for example efficient regression testing or the coordination of multiple stakeholders.

The wide range of variants identified through our case study indicates that contextual factors affect how TCR is applied. For this reason, we cannot provide any general guidelines for how to apply TCR for a specific case. However, our findings indicate that a successful application of TCR requires a combination of factors such as strong testing competence, a customer or product manager who is technical enough to read and understand test cases, adequate tool support for tracing requirements to automatic test cases, and a consistent implementation throughout the development life cycle. In addition, TCR seems to be less suitable to apply to quality requirements due to the challenges in defining good and efficient test cases for these complex requirements.

To summarize, the results provide empirically-based insights that can aid practitioners in improving their agile requirements practices. Furthermore, the facets used to characterize TCR variants can be used by practitioners to pinpoint a particular TCR variant suitable to their needs. The facets also provide a basis for further research. For example, future studies could investigate how requirements format and TCR-specific tool support can support the RE activities and thereby improve the coordination and effectiveness of agile development. This could involve document studies of the TCR artefacts for our case companies. Another interesting research direction could be to investigate factors, relationships and limits at play when introducing TCR and how these are affected by different contexts with the aim of providing more specific recommendations for which TCR variant to apply depending on, e.g. project size, requirements volatility, number and types of stakeholders involved. An interesting outcome of such research would be a TCR reference process that can be parameterized by contextual factors.

**Acknowledgement.** We want to thank all the participants in the interviews and the focus group sessions. This work was funded by EASE (ease.cs.lth.se).

# 8   Appendix

The following protocol was used to collect empirical data in the second iteration of study, see Section 4.2.

**Interviewee characteristics**
Role and length of experience in current role and total work experience

**Main project characteristics**
Number of project members, requirements & test cases, applied process model, project context and lead time etc.

**Information flow between requirements and testing**

- What artefacts are used, created/read/update (CRU)?
- For each artefact: what is its purpose, who creates, uses and changes it?
- What links or connections are there between the artefacts?
- How does the TCR practice (or the artefacts involved) support Lauesen's requirements roles (# is priority order during data collection):
   – Elicitation and validation of requirements (3)
   – Verification of requirements (1)
   – Requirements management: managing change including traceability (2)
   – Customer interaction & commitment / Contractual purposes (4)

- For each requirements role (RR):
   – How does TCR work in this RR? And, why or why not?
   – What are the benefits and challenges when using TCR in this RR? Why does TCR incur these?
   – What improvements/possibilities can you think of using TCR in this RR?